# NISE Estimation of an Economic Model of Crime

Eric Blankmeyer

Professor Emeritus of Economics
Department of Finance and Economics
Texas State University
San Marcos TX 78666
Email eb01@txstate.edu

March 2020

**Abstract**. An economic model of crime is used to explore the consistent estimation of a simultaneous linear equation without recourse to instrumental variables. A maximum-likelihood procedure (NISE) is introduced, and its results are compared to ordinary least squares and two-stage least squares. The paper is motivated by previous research on the crime model and by the well-known practical problem that valid instruments are frequently unavailable.

# NISE Estimation of an Economic Model of Crime

## 1. Introduction

The consistent estimation of a simultaneous linear equation is a challenging problem in practice. When identification is based on exclusion restrictions, the apparently-universal recommendation is to use instrumental variables (IV), for example two-stage least squares (TSLS). "The various methods that have been developed for simultaneous-equations models are all IV estimators" (Greene 2003, 398). However, the IV approach is frequently problematic since a valid instrument must be uncorrelated with the random component of the variable for which it is a surrogate but sufficiently correlated with its systematic component. "Those who use instrumental variables would do well to anticipate the inevitable barrage of questions about the appropriateness of their instruments" (Leamer 2010, 35). Concern about the effectiveness of IV methods is apparent in the literature on weak instruments, ably summarized by Hansen (2017, 316-323). When the instruments' exogeneity is in doubt, Conley et al. (2012) and Nevo and Rosen (2012) explore procedures to estimate bounds on the problematic regression coefficients.

This paper outlines the Non-Instrumental Simultaneous-Equation (NISE) estimator, which requires no instruments and is consistent under a set of assumptions often invoked for linear statistical models. Although it is asymptotically inefficient relative to IV estimators when valid instruments are at hand, NISE may be applicable in several situations: (i) observations on the instruments are unavailable or incomplete; (ii) the instruments are found to be weak; (iii) they fail Sargan's J test for exogeneity; or (iv) a researcher simply wants a second opinion about her IV estimates. The next

___________________

**Acknowledgement.** This paper is derived in part from an article published in *Applied Economics Letters* 25: 1097-1100.
DOI: 10.1080/13504851.2017.1397847.

section provides a concise description of the NISE estimator; analytical details, simulations and additional applications may be found in the papers by Blankmeyer (2013, 2018). In section 3, OLS, TSLS and NISE are applied to an economic model of crime. The last section is reserved for a few conclusions and caveats.

## 2. The NISE estimator

The simultaneous linear equation of interest is

$$\mathbf{Y}\boldsymbol{\gamma} = \mathbf{X}\boldsymbol{\beta} + \mathbf{u}, \tag{1}$$

where the matrix $\mathbf{Y}$ contains observations on G endogenous variables, $\mathbf{X}$ contains observations on H exogenous variables, $\boldsymbol{\gamma}$ and $\boldsymbol{\beta}$ are vectors of unknown parameters, and $\mathbf{u}$ is a vector of unobservable gaussian disturbances identically and independently distributed with $E(\mathbf{u}) = E(\mathbf{X}\mathbf{u}) = \mathbf{0}$. In addition to $\mathbf{X}$, the model contains L exogenous variables that appear in other linear equations; and L ≥ G so (1) is identifiable by exclusion restrictions. A researcher wants to estimate (1) but does not intend to estimate the model's other equations. In fact, she may have no usable data on the instruments, which is a principal motive for choosing NISE instead of an IV estimator. *It is irrelevant for identification that the actual measurements on the excluded exogenous variables may be absent from the data set or otherwise unusable.* In a correctly-specified model, the real effects of the excluded exogenous variables on the equation of interest are reflected in $\mathbf{Y}$.

For this linear system, only one of whose equations is to be estimated, Davidson and MacKinnon (1993, 644) show that there is no Jacobian term in the log likelihood; accordingly, the NISE maximum-likelihood estimator minimizes the Lagrangian

$$F = (\mathbf{Y}\boldsymbol{\gamma} - \mathbf{X}\boldsymbol{\beta})^T(\mathbf{Y}\boldsymbol{\gamma} - \mathbf{X}\boldsymbol{\beta}) - \lambda[\boldsymbol{\gamma}^T(\mathbf{Y}^T\mathbf{Y})\boldsymbol{\gamma} - 1], \tag{2}$$

where the constraint guarantees that the total squared error is minimized with respect to the left-hand side of equation (1). In an illuminating survey of the various simultaneous-equation estimators, Chow (1964, 533-537,

542-544) explains the relationships among IV, canonical correlation and the estimator that I call NISE. In fact equation (1) can be handled with standard software that computes the largest squared canonical correlation between **Y** and **X**: the canonical coefficients of **Y**, denoted **c**, estimate γ; and β is estimated by the OLS regression of **Yc** on **X**. A researcher may then choose to renormalize the equation, dividing both sides by an element of **c**. Finally, a pairs bootstrap will approximate the sampling errors of these NISE coefficients.

## 3. An economic model of crime

In this section OLS, TSLS and NISE are applied to an economic model of crime. The data set "Crime" (Croissant 2015) is a panel of 90 counties in North Carolina from 1981 to 1987 first examined by Cornwell and Trumbull (1994, hereafter denoted CT). Both CT and Baltagi (2006) use the data to assess, *inter alia*, how various law-enforcement measures affect the crime rate. An issue is the possibility of simultaneity bias: for example, more arrests may discourage criminal behavior, but it is also plausible that more crime results in more arrests. Likewise a larger police force may deter illegal activity but may also cause more offenses to be detected or reported; or a higher crime rate may lead to the recruitment of more police (Baltagi 2006, 544). The authors address this simultaneity issue with two instrumental variables. The first is the offense "mix, which is the ratio of crimes involving face-to-face contact (such as robbery, assault and rape) to those that do not. The rationale for using this variable is that arrest is facilitated by positive identification of the offender. The second instrument is per capita tax revenue. This is justified on the basis that counties with preferences for law enforcement will vote for higher taxes to fund a larger police force" (Baltagi 2006, 544).

The dependent variable is the annual per-capita crime rate in each county. The potential deterrents include the probability of arrest (arrests per crime), the probability of being convicted (convictions per arrest), and the probability of incarceration (imprisonments per conviction). The number of police per capita is a fourth potential deterrent. The authors propose that good employment opportunities –as measured by the average wage in the local manufacturing sector—should also be expected to reduce criminality. All these variables are log transformed and enter the linear regression as deviations from their county averages over the seven years. There are accordingly 630 observations. CT "include time effects to capture variations in the crime rate common to all counties. For convenience, we omit them from the formal presentation of our model" (1994, 362 footnote 3); and so do I.

Table A shows that, except for the police variable, all the regression coefficients have the expected negative signs. The coefficient for per-capita police is positive, suggesting that no net deterrent effect is associated with more law-enforcement personnel, *cet. par.* In each regression method, the coefficient for the arrest probability is larger in magnitude than the coefficient for the probability of conviction, which is in turn larger in magnitude than the coefficient for imprisonment probability. CT (1994, 361) remark that some versions of the economic model of crime predict this ranking.

Because Table A includes most of the salient regressors identified by CT(1994) and by Baltagi (2006), it is not surprising that the OLS coefficients are quite close to their counterparts in the "Fixed Effects" column of Table I in Baltagi (2006, 545). Likewise the TSLS coefficients are generally similar to the corresponding values in the FE2SLS column of his Table I. In Table A, the NISE coefficients for probability of arrest, conviction and imprisonment are larger in magnitude that their OLS and TSLS counterparts and larger than the corresponding estimates found by CT and by Baltagi. However, those NISE coefficients are well within the range of values estimated in previous research on crime-deterrence effects as reported in Table 1 of CT (1994, 362).

Table A shows that conventional levels of statistical significance are achieved by all the OLS regression coefficients and by all the NISE coefficients except the manufacturing wage rate, but no TSLS coefficient is significant. The latter finding is consistent with Baltagi's FE2SLS results.

OLS assumes that all five regressors are exogenous while NISE and TSLS allow for the possible endogeneity of the arrest rate and police per capita. With respect to TSLS, Sargan's J test cannot be used to assess the instruments' validity since the model is just identified. In lieu of a formal

Hausman endogeneity test, the last rows of Table A indicate that a bootstrap test comparing OLS and TSLS cannot reject the hypothesis that OLS is consistent, while a bootstrap test comparing OLS and NISE suggests that OLS is inconsistent.

Table A does not replicate all the results in CT (1994) and Baltagi (2006). Those authors examine additional regressors, econometric techniques and obstacles to consistent estimation. I have focused more narrowly on the role of NISE in a situation where the TSLS results are inconclusive.

## 4. Conclusions and caveats

When a linear model may be subject to simultaneity bias, NISE is proposed as an alternative (or a complement) to IV estimators. This paper has explored simultaneity bias in an economic model of crime previously examined by CT (1994) and Baltagi (2006). I find that the TSLS regression is hampered by large standard errors, but OLS and NISE appear to differ significantly with respect to the coefficients of the variables believed to be endogenous.

While these results tend to support the use of NISE for the estimation of a simultaneous linear equation, I note that NISE is not effective against bias in other situations where IV is frequently applied, e. g. "errors in variables" and "omitted variables." It is also important to recognize that IV is not the only identification strategy for a linear model with endogeneity; for example, restrictions on certain covariances can also be considered (e. g., Lewbel 2012).

Finally, a pairs bootstrap can produce standard errors for NISE; but it is preferable to apply the bootstrap to a more robust estimator of dispersion, e. g. the median absolute deviation or the Qn statistic (Rousseeuw and Croux 1993; Maronna et al. 2006, chapter 2). In addition to limiting the distortions due to outlying data points, a robust version of the standard error is desirable since the NISE coefficients do not necessarily have a finite second moment, as Anderson (2010) explains in the context of the limited-information maximum-likelihood estimator.

# References

Anderson, T., 2010. The LIML estimator has finite moments ! *Journal of Econometrics* 157: 359-361.

Baltagi, B. 2006. Estimating an economic model of crime using panel data from North Carolina. *Journal of Applied Econometrics* 21: 543-547.

Blankmeyer, E., 2013. Structural-equation estimation without instrumental variables. Social Science Research Network paper 2316436, available at http://www.ssrn.com/en/. Also available as arXiv paper 1709.09512 at https://arxiv.org.

Blankmeyer,E., 2018. Simultaneous-equation estimation with no instrumental variables. *Applied Economics Letters* 25: 1097-1100. DOI: 10.1080/13504851.2017.1397847.

Chow, G., 1964. A comparison of alternative estimators for simultaneous equations. *Econometrica* 32: 532-553.

Conley, T., C. Hansen, P. Rossi, 2012. Plausibly exogenous. *Review of Economics and Statistics* 94: 260-272.

Cornwell, C., W. Trumbull. 1994. Estimating the economic model of crime with panel data. *Review of Economics and Statistics* 76: 360-366.

Croissant, Y., 2015. *Ecdat: data sets for econometrics*. Available at https://cran.r-project.org/web/packages/.

Davidson, R., J. MacKinnon, 1993. *Estimation and Inference in Econometrics*. New York: Oxford University Press.

Greene, W., 2003. *Econometric Analysis*, fifth edition. Upper Saddle River NJ: Prentice Hall.

Hansen, B. 2017. *Econometrics*. Available at https://www.ssc.wisc.edu/~bhansen/econometrics/

Leamer, E., 2010. Tanatalus on the road to Asymptopia. *Journal of Economic Literature* 24: 31-46.

Lewbel, A., 2012. Using heteroskedasticity to identify and estimate mismeasured and endogenous regressor models, *Journal of Business and Economic Statistics* 30: 67-80.

Maronna, R., R. Martin, V. Yohai, 2006. *Robust Statistics: Theory and Methods*. Chichester, England: John Wiley & Sons.

Nevo, A., A.Rosen, 2012. Identification with imperfect instruments. *Review of Economics and Statistics* 94: 659-671.

Rousseeuw, P., C. Croux, 1993. Alternatives to the median absolute deviation. *Journal of the American Statistical Association* 88: 1273-1283.

## Table A. Estimates of the crime model

|  | OLS | NISE | TSLS |
|---|---|---|---|
| ln probability of arrest | -0.359<br>0.030 | -1.140<br>0.105 | -0.611<br>0.489 |
| ln probability of conviction | -0.285<br>0.019 | -0.689<br>0.063 | -0.442<br>0.303 |
| ln probability of prison | -0.176<br>0.030 | -0.428<br>0.090 | -0.263<br>0.172 |
| ln police per capita | 0.418<br>0.024 | 0.938<br>0.085 | 0.679<br>0.507 |
| ln manufacturing wage | -0.327<br>0.100 | -0.181<br>0.238 | -0.244<br>0.196 |

|  | NISE - OLS | TSLS-NISE | TSLS-OLS |
|---|---|---|---|
| ln probability of arrest | -0.781<br>0.126 | 0.529<br>0.603 | -0.252<br>0.561 |
| ln police per capita | 0.520<br>0.087 | -0.259<br>0.626 | 0.261<br>0.606 |

Note: the dependent variable is ln crime rate, n = 630; estimated standard errors are listed below their respective regression coefficients; the NISE standard errors are computed using a pairs bootstrap and the robust Qn statistic of Rousseeuw and Croux (1993).